\documentclass[12pt]{article}
\usepackage{epsfig}
\usepackage{fullpage}
\usepackage[psamsfonts]{eucal}
\usepackage{amsfonts}
\usepackage{amssymb}
\begin{document}

\begin{titlepage}
\vspace{-1mm}

\begin{flushright}
 M\'exico ICN-UNAM\\
 \LaTeX
\end{flushright}
\vskip 20mm

\vspace{1mm}
\begin{center}{\bf\Large SIMPLE EVALUATION OF FRANCK-CONDON FACTORS
    AND NON-CONDON EFFECTS IN THE MORSE POTENTIAL}
\end{center}
\vskip 12mm

\begin{center}
  {\bf\large 
    J.C.~ L\'opez V.
    ${}^{\dag}${\normalsize\footnote{vieyra@nuclecu.unam.mx}\vspace{8mm}}, 
    A.L.~ Rivera
    ${}^{\ddagger}${\normalsize\footnote{rivera@ciicap.uaem.mx}\vspace{8mm}},
    Yu.F.~Smirnov${}^{\dag}${\normalsize\footnote{smirnov@nuclecu.unam.mx}}
    and 
    A.~Frank${}^{\dag}{}^\S$\footnote{frank@nuclecu.unam.mx}}\\ 
    {\em${}^{\dag}$ Instituto de Ciencias Nucleares, UNAM, \\
     Apartado Postal 70-543,    04510 Mexico D.F., M\'exico \\[10pt]
    ${}^{\ddagger}$ Centro de Investigaci\'on en Ingenier\'{\i}a y
    Ciencias Aplicadas, UAEM\\
    Apartado Postal 6-78, 62131 Cuernavaca Morelos, M\'exico\\[10pt]
    ${}^\S$Centro de Ciencias F\'{\i}sicas, UNAM,\\
    Apartado Postal 139-B Cuernavaca Morelos, M\'exico }

\vskip 2cm

{\Large Abstract}
\end{center}

\vskip 0.5 cm

\begin{quote}
  The calculation of Franck-Condon factors between different
  one-dimensional Morse potential eigenstates using a formula derived
  from the Wigner function is discussed. Our numerical
  calculations using a very simple program written in {\it Mathematica}
  is compared with other calculations. We show that our results have
  a similar accuracy as the calculations performed with more
  sophisticated methods. We discuss the extension of our method to
  include non-Condon effects in the calculation.
\end{quote}
\vskip 1 cm
\end{titlepage}


\section{Introduction}

The study of complex features of molecular vibrational spectra at
high overtones and combinations has become possible by recent
advances in optical laser spectroscopy. Rovibrational level
structure of small polyatomic molecules has been elucidated in a
formerly inaccessible energy range. Transition intensities between
these levels can reveal fine aspects of the structure and
dynamical properties of these systems. Theoretically, the
evaluation of molecular vibrational wave functions has relied upon
a number of assumptions and approximations, of which the use of a
harmonic normal-mode basis is the best known procedure. At higher
energies, however, normal-mode vibrational assignments of
individual states becomes more difficult due to anharmonic
effects. In this context, Morse potentials \cite{Morse:1929} have
been proposed to more accurately model molecular systems, as they
often constitute good approximations to diatomic molecular
potentials derived {\it ab
  initio} \cite{Child:84}. Even if the general solutions to the
corresponding Schr\"odinger equation cannot be written in analytical
form, the case of zero angular momentum can be closely approximated by
the one-dimensional analytical solutions \cite{Franck:AM}.  The
analytic form of these wave functions has been used not only for the
simple diatomic case, but for the analysis of vibrational structure in
polyatomic molecules, within a local-mode framework (see eg.
\cite{Iach:JCP98,Muller:98,Muller:99,Muller:2000}). In addition, the
connection of the one-dimensional Morse, Poeschl-Teller and other
exactly solvable potentials with the $SU(2)$ group has been exploited
in the formulation of algebraic models which incorporate the
anharmonic character of the local vibrations from the outset (see
\cite{Frank:99} and references therein).

In recent studies, the Lie-algebraic approach has been also applied to
the evaluation of Franck-Condon intensities \cite{Franck:25,
  Condon:1926} (FC) in small polyatomic molecules
\cite{Muller:98,Muller:99,Muller:2000}. These studies accurately
reproduce the emission spectra of the $S_2O$\, ${\tilde C}^1 A' -
{\tilde X}^1 A'$ $(\pi^\star - \pi)$ experimental intensities
\cite{Muller:99}, requiring for their calculation a combination of two
procedures: a) an accurate fit of the vibrational energy levels in
both electronic states using a local-mode algebraic Hamiltonian, and
b) an efficient and accurate methodology for the evaluation of
one-dimensional Morse wave function overlaps. One possibility to
calculate overlaps of Morse wave functions is to use an approximation
based on an energy-dependent harmonic approximation where the widths
and displacement of the harmonic potentials are adjusted in order to
mimic the behavior of the Morse potentials at the corresponding energy
levels \cite{Iach:JCP98}. Although this method gives explicit formulas
for the overlap integrals, and consequently a good accuracy for the
low lying vibrational states, it departs from the exact results at
higher energies. Other (numerical) methods, like the use of the
Configuration Localized States (CLS) formalism and the use of Gaussian
quadratures \cite{Carvajal:99,curro:99} have been proposed as
alternative ways to evaluate the overlaps of two Morse functions.

In view of future applications to the evaluation of
multidimensional Franck-Condon factors, a simple and reliable
method of calculation of one-dimensional Morse functions overlaps
has to be devised. In this paper we propose a new method of
evaluation based on our recent work on the Wigner function of the
Morse potential eigenstates \cite{Frank:PRA2000}.  Our main result
contains analytical sums of simple one-dimensional integrals which
are not tabulated and which must be evaluated with standard
numerical techniques. We make no approximations but  must rely on
numerical evaluations for part of our calculations. Fortunately,
present computational resources (like {\it Mathematica}) can
handle these calculations with very high accuracy in a  simple and
efficient way.  We show that a short program written in {\it
Mathematica} (included in an appendix) gives accurate results for
several molecular examples.  To demonstrate the reliability of the
method  we compare our results with other calculations and provide
estimates for its range of applicability. In the last section we
show that the procedure can be extended, with  no loss of
accuracy, in order to include non-Condon effects.\\

\section{The Morse Potential Eigenstates}
\vskip 0.2cm

In what follows we briefly review the properties of the Morse
potential eigenstates and the use of the corresponding Wigner
function to evaluate their overlap. We then express our result in
terms of the integral ${\cal I}_a$ (eq. (\ref{Ia})), for which we
find a symmetry property and a recurrence relation, which can be
used to verify the accuracy of its numerical evaluation. After
discussing some particular analytic solutions, we apply our method
to several examples and compare our results with more sofisticated
techniques \cite{Muller:99,Carvajal:99,Ley-Koo:95}.\\

\vskip 0.2cm

The Morse potential is characterized by two parameters: its depth
${\cal D}$ (bonding energy), and the inverse of its range $\beta
>0$ (restitution constant in the harmonic approximation). In the
one-dimensional version it is given by

\begin{equation}
\label{VMorse}
V(x) = {\cal D} \bigg[ ( 1 - e^{-\beta x} )^2 - 1 \bigg] \, , \quad
(-\infty < x < \infty),
\end{equation}
where $x$ is the position with respect to the equilibrium point.

It is well known that the discrete spectrum of the Morse potential
is described by the formula

\begin{equation}
\label{eigval}
 E_\nu = - \frac{\hbar^2 \beta^2}{2\mu} (j-\nu)^2 \, ,
\end{equation}

where

\begin{equation}
\label{jrel}
j = \frac{\sqrt{2\mu {\cal D}}}{\beta\hbar} - \frac{1}{2}
\end{equation}
while $\nu$ is the number of anharmonic phonons $\nu=0,1,\ldots
\lfloor j \rfloor$, with  $\lfloor j \rfloor$ being the largest
integer not exceeding $j$, and $\mu$  the reduced mass of the
molecule.

If $j$ takes an integer or half integer value, it determines the
irreducible representation $\mathfrak{D}^{j}$ of the $SU(2)$
algebra which is the dynamical algebra of the Morse potential. In
general there are $\lfloor j \rfloor + 1$ bound states, except
when $j$ is an integer, in which case  one state has zero energy.

The eigenfunctions corresponding to the eigenvalue (\ref{eigval})
are given by

\begin{equation}
\label{eigfun}
 \psi_{j,\nu}^\beta (\xi) = {\cal N}_{j,\nu} e^{-\xi/2}
\xi^{j - \nu} L_\nu^{2(j-\nu)}( \xi)\, ,
\end{equation}
where a new variable (the Morse coordinate)

\begin{equation}
 \label{Morsecoord}
 \xi = (2j+1) e^{-\beta x} \, ,
\end{equation}
has been  introduced.  The wave function (\ref{eigfun}) includes
the associated Laguerre polynomials \cite{Gradshteyn}

\begin{equation}
\label{Laguerre}
L^\alpha_{\nu} (\xi) = \sum^\nu_{k = 0} ( - 1 )^k { \Gamma (\nu +
\alpha + 1) \over k!(\nu-k)! \Gamma (k + \alpha+1)} \xi^k ~~ ,
\end{equation}
normalized by the condition

\begin{equation}
\int^\infty_{-\infty} { dx | \psi_{j \nu}^\beta (x) |^2}= 1 \, ,
\end{equation}
when the normalization factor is taken in the form
\begin{equation}
\label{normalization}
 \big({\cal N}_{j \nu}^\beta \big)^2 = {\beta
\nu! 2 (j-\nu) \over \Gamma (2j - \nu+1)} ~~ .
\end{equation}
The eigenvalues (\ref{eigval}) can be written as
\begin{equation}
E_\nu = - {\cal D} + \hbar \omega_e (\nu + 1/2) - \hbar
\omega_e\chi (\nu + 1/2)^2 \, ,
\end{equation}
where
\begin{eqnarray}
 \hbar\omega_e = \beta \hbar \sqrt{{2 {\cal D} \over \mu}} ~~
,  \\
 \hbar \omega_e \chi = {\hbar^2 \beta^2 \over 2\mu} ~~ . 
\end{eqnarray}
These relations relate the parameters of the Morse potential with
the standard spectroscopic parameters $\omega_e$ and $\omega_e
\chi$ \cite{Herzberg}. In particular we have
\begin{equation}
2j+1 = {\omega_e \over \omega_e \chi} = 4 {{\cal D} \over \hbar
\omega_e} ~~ ,
\end{equation}
and
\begin{equation}
  \beta =  \sqrt{\frac{4\pi c}{\hbar} \mu\,\omega_e \chi} \ \, .
\end{equation}

\section{Franck-Condon factors for the Morse Potential}
\vskip 0.2cm

Molecular Franck-Condon factors describe, in first approximation, the
dependence on the vibrational wave functions of transition intensities
between vibrational states in different electronic configurations.
Assuming that the electronic wave functions are independent of the
vibrational states (Condon approximation \cite{Condon:1926}), the
Franck-Condon factors are given by the square of the overlap integral
between initial and final wave functions in the transition
\begin{equation}
\label{FCdef}
  F_{1,2} \equiv \vert f_{1,2} \vert^2
\end{equation}
where
\begin{equation}
f_{1,2} =
\langle
 \psi_{j_1,\nu_1}^{\beta_1} (x - R_1) \vert
 \psi_{j_2,\nu_2}^{\beta_2} (x - R_2)
\rangle\, ,
\end{equation}
and $R_1$ and $R_2$ are  the equilibrium distances of the nuclei
in the initial and final electronic states, respectively. Relative
transition intensities, in the Condon approximation, are simply
related to the Franck-Condon factors (\ref{FCdef}) by
\cite{Muller:99}

\[
I_{1,2}\propto \nu^4 |f_{1,2}|^2 \, ,
\]
where the frequency dependent factor $\nu^4$ ($\nu$ is essentially the
difference between vibrational energies of the initial and final
states in the transition), is introduced for calibration with
experimental data.

In a higher approximation (Non-Condon), the electric dipole
operator for transitions is expressed as a series expansion on the
interatomic coordinates \cite{Renato:2001} or, equivalently, as a
series expansion in one of the Morse coordinates ${\xi}_1(x)$ or $
{\xi}_2(x)$ \footnote{These coordinates are the natural
coordinates appearing in
  the eigenfunctions of the Morse potential and are defined as
   ${\xi}_1(x)=(2j_1+1)\exp{(-\beta_1 x_1)}$ and ${\xi}_2(x)=
   (2j_2+1)\exp{(-\beta_2 x_2)}$ (see Eq. (\ref{Morsecoord}))}.
    Our method can be adapted to include such effects
    in a straightforward manner (see discussion below).

 A simple way to calculate the FC factors arises from the Wigner function of
two Morse wave functions, which we define in the form
\cite{Frank:PRA2000}

\begin{equation}
\label{Wigner}
  W( \psi_{j_1,\nu_1}^{\beta_1}, \psi_{j_2,\nu_2}^{\beta_2} \vert x,p )
= \frac{1}{2\pi\hbar} \int_{-\infty}^{\infty}
dr\, {\psi_{j_1,\nu_1}^{\beta_1}}^* (x-\frac{1}{2}r) \, e^{-ipr/\hbar}
\, \psi_{j_2,\nu_2}^{\beta_2} (x+\frac{1}{2}r) \, .
\end{equation}

Making the change of variable $z=\frac{1}{2} r$ and taking into
account that

\begin{equation}
  {\psi_{j_1,\nu_1}^{-\beta_1}} (x-z) =
\left( \frac{{\bar{\cal N}}_{j_1 \nu_1}^{(\beta_1)}}
      {{\cal N}_{j_1 \nu_1}^{\beta_1}}
 \right)
{\psi_{j_1,\nu_1}^{\beta_1}} (z-x) =
\psi_{j_1,\nu_1}^{\beta_1}(z-x) \, ,
\end{equation}
where ${\psi_{j_1,\nu_1}^{-\beta_1}} (x)$ is the wave fuction
corresponding to the mirror image of the Morse potential
(\ref{VMorse}) whose normalization ${\bar{\cal N}}_{j_1
\nu_1}^{(+\beta_1)}$ is clearly also given by
(\ref{normalization}).

We then find the expression
\begin{equation}
  W( \psi_{j_1,\nu_1}^{-\beta_1}, \psi_{j_2,\nu_2}^{\beta_2} \vert x,p )
= -\frac{1}{\pi\hbar} \int_{-\infty}^{\infty}
dz\, {\psi_{j_1,\nu_1}^{\beta_1}}^* (x-z) \, e^{-2 i pz/\hbar}
\, \psi_{j_2,\nu_2}^{\beta_2} (z+x) \, .
\end{equation}

If we evaluate the integral at $(x=\frac{1}{2}(R_2 - R_2), p=0)$ we
get precisely the overlap integral between
$\psi_{j_1,\nu_1}^{\beta_1}$ and $\psi_{j_2,\nu_2}^{\beta_2}$, and
thus

\begin{equation}
  f_{1,2} = -\pi \hbar
W(\psi_{j_1,\nu_1}^{-\beta_1}, \psi_{j_2,\nu_2}^{\beta_2} \vert
x=\frac{1}{2}(R_2 - R_2),p=0) \, .
\end{equation}

In Appendix A we calculate the Wigner function corresponding to
two different Morse potential states (\ref{eigfun}), leading to
the following expression for the Franck-Condon factor:

\begin{eqnarray}
\label{f12}
 f_{1,2} &=& 2
\sqrt{
{ a \nu_1 ! \nu_2 !  (j_1 - \nu_1) (j_2 -\nu_2)
\over
\Gamma (2 j_1 - \nu_1 + 1) \Gamma (2 j_2 - \nu_2 +1)}
} {y_1}^{j_1-\nu_1} {y_2}^{j_2-\nu_2} \,
 \nonumber \\[10pt] &&
\times  \sum^{\nu_1, \nu_2}_{{\ell}_1 , {\ell}_2 = 0 }
{( - 1)^{{\ell}_1 + {\ell}_2}  \over {\ell}_1! {\ell}_2 !}
\left(
\begin{array}{c}
2j_1-\nu_1 \\ \nu_1 - {\ell}_1
\end{array}
\right)
\left(
\begin{array}{c}
2j_2 - \nu_2 \\ \nu_2 -{\ell}_2
\end{array}
\right) \, {y_1}^{\ell_1} {y_2}^{\ell_2} \,
\nonumber \\[10pt] & &
\times \  {\cal I}_a ( j_1 - \nu_1 + \ell_1,  j_2 - \nu_2 + \ell_2, y_1,y_2)
~~ ,
\end{eqnarray}
where
\begin{eqnarray}
  y_1 = (2 j_1 +1) e^{-\beta_1 (R_2-R_1)/2} \, ,\nonumber \\
  y_2 = (2 j_2 +1) e^{+\beta_2 (R_2-R_1)/2} \, ,
\end{eqnarray}
and
\begin{equation}
\label{Ia}
{\cal I}_a (k_1, k_2, y_1, y_2) = \int^\infty_0 du \,\, u^{k_1 + a k_2 - 1}
e^{\displaystyle -\frac{1}{2} (y_1 u + y_2 u^a) } \, ,
\end{equation}
with $a=\frac{\beta_2}{\beta_1}$, $k_1=j_1 - \nu_1 + \ell_1$ and
$k_2 = j_2 - \nu_2 + \ell_2$.

Overlap integrals of the type (\ref{Ia}) were previously
encountered in ref.\cite{Fraser:53}, where it is stated that they
``cannot be evaluated exactly in general (for $a\ne 1$)'' and ``it is
necessary to resort to a valid approximation''. In addition, it
was stated that the alternating sums involved in (\ref{f12}) may
give unstable results as a consequence of the numerical error
accumulation when the sums are performed over very large numbers,
a  situation that can arise for  realistic calculations involving
large values of $j_1$ or $j_2$.

As mentioned in the introduction, several approaches have been
proposed to find FC factors (\ref{f12}) in analytic form,
including the use of different versions of the harmonic oscillator
approximation of the Morse wave functions \cite{Iach:JCP98,
Muller:98,Muller:99,
  Berrondo:80,Rivas:92}, the expansion of the integral (\ref{Ia}) into
series containing polygamma functions \cite{Matsumoto:93}, the use
of a Laguerre quadrature \cite{Carvajal:99}, and others. We now
show that the integral (\ref{Ia}), although not tabulated,  is an
analytical well-behaved function which satisfies  relations which
can be used for its reliable evaluation using a personal computer.
In particular we have found that the program {\it Mathematica} can
be applied to calculate those integrals with high accuracy in an
efficient way.\\

We start our study by stating the following properties of the integral
${\cal I}_a$ :

\begin{enumerate}
\item Symmetry property:
\begin{equation}
\label{symmetry}
{\cal I}_a (k_1, k_2, y_1, y_2) = {1\over a} \ {\cal I}_{1/a}
(k_2, k_1, y_2, y_1)
 ~~ ,
\end{equation}
which follows from (\ref{Ia}) and the change of variables
$u^\prime=u^a$.

As an application of (\ref{symmetry}) we can check the accuracy of the
integration appearing in the transition $A^1 \Sigma^+_u(\nu=0) \ - \ 
X^1 \Sigma^+_g (\nu=0)$ of the $^7$Li$_2$ molecule (Table
\ref{villa:T5}).  In this case the difference of both calculations
${\cal I}_a - {1\over a} \ {\cal I}_{1/a}$ is $-1.71551\times
10^{-67}$ where ${\cal I}_a \approx {1\over a} \ {\cal I}_{1/a}
\approx 1.14752\ldots\times 10^{-53}$. This shows the typical
precision in the computations of integrals (\ref{Ia}). In order to get
such an accuracy we have used a scaling method for the integral. This
procedure is discussed below.

\item Recurrence relation:
\begin{eqnarray}
\label{recurrence} {\cal I}_a (k_1, k_2, y_1, y_2) &=& {1\over
y_1} [ 2(k_1 + a k_2 -1){\cal I}_a (k_1 - 1, k_2, y_1, y_2)
\nonumber \\ && - a y_2 {\cal I}_a (k_1 - 1, k_2 + 1, y_1, y_2) +
\delta_{\scriptstyle\alpha,1}] ~~ \, ,
\end{eqnarray}
where $\alpha = k_1 + a k_2$. This property follows by carrying out
the integration by parts. Note that $\alpha=1$ can only occur for the
case $a=1$ ($\beta_1=\beta_2$).
\end{enumerate}

Regarding the application of the three-term recurrence relation
(\ref{recurrence}), we observe that it can be used to find one such
integral if the other two included in this relation are known. It is
therefore possible to find a full set of integrals (\ref{Ia}) with all
possible values of $k_1=1,\ldots N_1, k_2=1,\ldots N_2$ if the
integrals ${\cal I}_a(k_1=1\ldots N_1+1,k_2=1,y_1,y_2) $ and ${\cal
  I}_a(k_1=1, k_2=1\ldots N_2+1,y_1,y_2)$ are known for given $a,
y_1$, and $y_2$. The relation (\ref{recurrence}) can be used in two
ways: to check the accuracy of the numerical integrations and to
calculate the values of the other
integrals.\\

As example we use the recurrence relation (\ref{recurrence}) to check
the accuracy of the integrations involved in the transition $A^1
\Sigma^+_u(\nu=6) \ - \ X^1 \Sigma^+_g (\nu=6)$ of $^7$Li$_2$ molecule
(see Table \ref{villa:T5}). Let us consider the integral ${\cal I}_a$
(Eq. \ref{Ia}) with $(\ell_1=0,\ell_2=1)$ and compute it by numerical
integration and with the recurrence relation (\ref{recurrence}). For
this we need to compute integrals with $(\ell_1=0,\ell_2=0)$ and
$(\ell_1=0,\ell_2=1)$. The absolute difference of both calculations is
$3.17601\times 10^{-67}$, being ${\cal I}_a \approx
2.07527\ldots\times 10^{-53}$, in agreement with the estimated
precision obtained with the symmetry relation. \\

Relations (\ref{symmetry}) and (\ref{recurrence}) led to a simple
method to verify the accuracy of the numerical integration. We are
currently studying other properties of the function (\ref{Ia}),
which is a generalization of the {$ \Gamma$} function, to which it
reduces in the case $a=1$.\\

In realistic cases, quite often the overlaps must be calculated for
states with rather large values of the parameters $j_1$ or $j_2$ (or
equivalently for large bonding energies, see Eq. (\ref{jrel}). In this
cases the numerical integration of (\ref{Ia}) requires a special
treatment in order to achieve high accuracy.  Although the integrand
is a well behaved function of $x$ having a single maximum and going to
zero at $x\to 0$ and $x\to\infty$, the resulting integral, being a
very small number, could be difficult to evaluate due to the limited
precision of the numerical calculations. One way to deal with such
complications is to perform a change of variables in order to scale
the integrand in such a way that the integrand at the maximum acquires
a unit value at $x=1$. In all cases studied, we have found that the
best procedure to evaluate (\ref{Ia}) is by carrying out a scaled
integration. Our tests showed that the relative differences of the
integrals calculated using formula (\ref{Ia}) and the symmetry
relation (\ref{symmetry}) are very small ($\lesssim
10^{-13}$) when the scaled integrations are used in both cases. \\

As our first example we have calculated Franck-Condon factors for the
$S-S$ mode in the $S_2O$ molecule (Table \ref{tab:curroT3}). We found
a maximal relative difference $\simeq 3.14\times 10^{-14}$ between the
scaled integrals (\ref{Ia}) and the scaled integration using the
symmetry relation (\ref{symmetry}). Despite  the fact that the
numerical integration can be calculated with high accuracy, it is not
always enough to compensate for the loss of accuracy due to the
alternating sums involved in (\ref{f12}). This becomes more apparent
in the cases in which we have a significant number of terms in the
summations.  We can see this effect in the results of Table
\ref{tab:curroT3} (e.g.  for $\nu_{\widetilde X} = 7$ and
$\nu_{\widetilde C} = 6$).  The use of logarithms to evaluate the
products (and divisions) of the very large numbers appearing in
(\ref{f12}) is usually a good strategy to avoid some problems. However
we have also calculated the FC factors using this strategy and found
no significant improvement in the results.

In Table \ref{tab:curroT3} we compare our results with calculations
using the method of quadratures proposed in \cite{Carvajal:99,curro:99}, and
with the algebraic formulation using modified harmonic functions of
ref. \cite{Iach:JCP98}. Our results are in excellent agreement with
the results of ref. \cite{Carvajal:99} except for the corner in the
table close to $\nu_{\widetilde X} =7$ and $\nu_{\widetilde C} =6$
where the calculations exhibit some deviations due to the above
mentioned loss of accuracy in the sums.

The approach using modified harmonic functions is in this case
expected to be a good approximation (at least for the overlaps
between the lowest-lying states) since we have large values of the
parameters $j_1, j_2$ (large depths ${\cal D}_1, {\cal D}_2$) .
However, there are significant deviations with respect to our
results even for this extreme case.\\

\bigskip
\noindent
For the sake of completeness we now discuss some  exactly solvable
examples of the FC formula (\ref{f12}): \\

\noindent
1. $\beta_1 = \beta_2 \quad (a = 1) $.
In this case we have \cite{Carvajal:99}:

\begin{eqnarray}
\label{Iaa1}
 {\cal I}_{a=1} (j_1 -\nu_1 +\ell_1,j_2 -\nu_2 +\ell_2,y_1,y_2 )
&=& \int_0^{\infty} du \exp{\left[ -\frac{1}{2} (y_1+y_2)u \right]}
\, u^{\alpha-1}
\nonumber \\
&=& \Gamma(\alpha) \left( \frac{y_1+y_2}{2} \right)^{-\alpha}
\end{eqnarray}
with $\alpha = k_1+k_2 = j_1 -\nu_1 +\ell_1 + j_2 -\nu_2 +\ell_2$, and thus
\begin{eqnarray}
\label{f12a1}
 f_{1,2}^{a=1} &=&
2 \sqrt{
{ \nu_1 ! \nu_2 !  (j_1 - \nu_1) (j_2 -\nu_2) \over
\Gamma (2 j_1 - \nu_1 + 1) \Gamma (2 j_2 - \nu_2 +1)}}
{y_1}^{j_1-\nu_1} {y_2}^{j_2-\nu_2} \,
\nonumber \\[10pt] &&
\times \sum^{\nu_1, \nu_2}_{{\ell}_1 , {\ell}_2 = 0 }
{( - 1)^{{\ell}_1 + {\ell}_2}  \over {\ell}_1! {\ell}_2 !}
\left(
\begin{array}{c}
2j_1-\nu_1 \\ \nu_1 - {\ell}_1
\end{array}
\right)
\left(
\begin{array}{c}
2j_2 - \nu_2 \\ \nu_2 -{\ell}_2
\end{array}
\right) \,  \\[10pt] & &
\times {y_1}^{\ell_1} {y_2}^{\ell_2} \,
 \left( \frac{2}{y_1+y_2}\right)^\alpha \Gamma(\alpha)
~~ . \nonumber
\end{eqnarray}

\noindent
If we define:
\begin{equation}
  \zeta = \frac{2j_2+1}{2j_1+1} e^{-\beta(R_2-R_1)} = \frac{y_2}{y_1}
  \, ,
\end{equation}
we obtain the same expression as Eq. (4.5) in Ref. \cite{Carvajal:99}.
The results for the Franck-Condon factors obtained trough the
numerical integration of (\ref{Ia}) and with the analytic formula
(\ref{f12a1}) are of comparable accuracy. This gives another
indication that the main source for the loss of accuracy is due to the
alternating sums appearing in (\ref{f12}), (\ref{f12a1}). To check the
accuracy of the numerical integration (\ref{Ia}), we have calculated
its relative difference with the analytical result (\ref{Iaa1}) for a
broad range of values (e.g. for $y_1=y_2=100$ and $\alpha\equiv
k_1+k_2 = 1\ldots 200$ ) and found a maximal relative difference of
$\Delta({\cal I}_a)\lesssim 5\times 10^{-13}$.  However we should note that
the analytical result for the integration (\ref{Iaa1}) can also lead
to uncertainties when it is numerically evaluated, specially for large
values of $\alpha$ and/or $y_1, y_2$.

\vspace{0.5cm}

\noindent
2. $\beta_2 = 2 \beta_1, \quad (a=2)$. In this case we have
\begin{eqnarray}
   {\cal I}_{a=2} (j_1 -\nu_1 +\ell_1,j_2 -\nu_2 +\ell_2,y_1,y_2 )
&=& \int_0^{\infty} du \exp{\left[ -\frac{1}{2} (y_1 u +y_2 u^2) \right]}
\, u^{\alpha-1}
\nonumber \\
&& \hspace{-100pt} = \bigg(y_2\bigg)^{\displaystyle -\frac{\alpha}{2}}
\, \Gamma(\alpha)
\exp{\left( \frac{y_1^2}{16 y_2}\right)}
D_{-\alpha}\left( \frac{y_1}{2\sqrt{y_2}} \right)\, ,
\end{eqnarray}
$D_{-\alpha}$ being the cylindrical functions \cite{Gradshteyn} with
$\alpha = j_1 -\nu_1 +\ell_1 - 2j_2 -2\nu_2 -2\ell_2$, and thus

\begin{eqnarray}
 f_{1,2}^{a=2} &=&
2 \sqrt{
{ 2 \nu_1 ! \nu_2 !  (j_1 - \nu_1) (j_2 -\nu_2) \over
\Gamma (2 j_1 - \nu_1 + 1) \Gamma (2 j_2 - \nu_2 +1)}}
{y_1}^{j_1-\nu_1} {y_2}^{j_2-\nu_2} \,
\nonumber \\[10pt] &&
 \sum^{\nu_1, \nu_2}_{{\ell}_1 , {\ell}_2 = 0 }
{( - 1)^{{\ell}_1 + {\ell}_2}  \over {\ell}_1! {\ell}_2 !}
\left(
\begin{array}{c}
2j_1-\nu_1 \\ \nu_1 - {\ell}_1
\end{array}
\right)
\left(
\begin{array}{c}
2j_2 - \nu_2 \\ \nu_2 -{\ell}_2
\end{array}
\right) \, \nonumber \\[10pt] & &
{y_1}^{\ell_1} {y_2}^{\ell_2} \,
\bigg(y_2\bigg)^{\displaystyle -\frac{\alpha}{2}}
\, \Gamma(\alpha)
\exp{\left( \frac{y_1^2}{16 y_2}\right)}
D_{-\alpha}\left( \frac{y_1}{2\sqrt{y_2}} \right)\,
~~ .
\end{eqnarray}

\vspace{1cm}
\noindent

3. $\beta_2 = \frac{1}{2} \beta_1$, $a=\frac{1}{2}$\\
In this case we have
\begin{eqnarray}
   {\cal I}_{a=\frac{1}{2}} (j_1 -\nu_1 +\ell_1,j_2 -\nu_2 +\ell_2,y_1,y_2 )
&=& \int_0^{\infty} du \exp{\left[ -\frac{1}{2} (y_1 u + y_2 \sqrt{u}) \right]}
\, u^{\alpha-1}
\nonumber \\[5pt]
&&\hspace{-100pt}
= 2 \, {\cal I}_2 (j_2 -\nu_2 +\ell_2,j_1 -\nu_1 +\ell_1,y_1,y_2) \, ,
\end{eqnarray}
where we have used the symmetry property (\ref{symmetry}), and therefore
\begin{eqnarray}
 f_{1,2}^{a=\frac{1}{2}} &=&
4 \sqrt{
{ 2 \nu_1 ! \nu_2 !  (j_1 - \nu_1) (j_2 -\nu_2) \over
\Gamma (2 j_1 - \nu_1 + 1) \Gamma (2 j_2 - \nu_2 +1)}}
{y_1}^{j_1-\nu_1} {y_2}^{j_2-\nu_2} \,
\nonumber \\[10pt] &&
 \sum^{\nu_1, \nu_2}_{{\ell}_1 , {\ell}_2 = 0 }
{( - 1)^{{\ell}_1 + {\ell}_2}  \over {\ell}_1! {\ell}_2 !}
\left(
\begin{array}{c}
2j_1-\nu_1 \\ \nu_1 - {\ell}_1
\end{array}
\right)
\left(
\begin{array}{c}
2j_2 - \nu_2 \\ \nu_2 -{\ell}_2
\end{array}
\right) \, \nonumber \\[10pt] & &
{y_1}^{\ell_1} {y_2}^{\ell_2} \,
\bigg(y_2\bigg)^{\displaystyle -\frac{\alpha}{2}}
\, \Gamma(\alpha)
\exp{\left( \frac{y_1^2}{16 y_2}\right)}
D_{-\alpha}\left( \frac{y_1}{2\sqrt{y_2}} \right)\,
~~ \, ,
\end{eqnarray}
with $\alpha = 2j_1 -2\nu_1 + 2\ell_1 +j_2 -\nu_2 +\ell_2$.

\bigskip

These particular cases can also be used to check the accuracy of the
general result (\ref{f12}, \ref{Ia}). \\

\section{Examples}

Continuing with our comparison with other approaches, we have
considered (i) the case of two identical (but displaced) Morse
potentials and (ii) the case of two different Morse potentials. We
have taken these examples from ref.  \cite{Carvajal:99}. Our results
are identical to those obtained in \cite{Carvajal:99} where an
analytical approximation for the Morse functions overlaps based on the
use an integration by quadratures was used \footnote{ In Ref.
  \cite{Carvajal:99} it was suggested to calculate the Franck-Condon
  matrix elements using the Laguerre quadrature in order to avoid the
  calculation of the alternative sum in (\ref{f12}).  However this
  approach may present a similar disadvantage since the calculation
  involves a sum over all positive and negative nodes of a
  polynomial.}  (see Tables (\ref{carvaj:TI}) and (\ref{carvaj:TII})).
Perhaps this is not surprising since the corresponding values of
$j_1,j_2$ are small. The accuracy of the numerical calculations is
high given that the effects of the numerical instabilities coming from
the alternating sums are relatively small.  The relevant point, as
observed in \cite{Carvajal:99}, is that the deviations in the results
obtained from harmonic functions relative to those using the Morse
functions are larger when the
depths ${\cal D}$ (and correspondingly the $j$ values) are small.\\


As a next set of examples we have considered the vibrational
transitions studied by Ley-Koo and collaborators \cite{Ley-Koo:95}. In
ref. \cite{Ley-Koo:95} Franck-Condon factors of diatomic molecules
were calculated by means of 3-dimensional Morse functions via the
method of confinement in a box.  The solutions of the radial
Schr\"odinger equation were found by diagonalization of the
Hamiltonian in a large basis ($N\sim 600$) of free-particle
eigenfunctions with appropriate boundary conditions.  We have borrowed
some tables from ref. \cite{Ley-Koo:95} to facilitate the comparisons.
Our results for the Franck-Condon factors are in excellent agreement
with those of Ley-Koo and collaborators (see Tables \ref{villa:T5},
\ref{villa:T6}, \ref{villa:T12}, \ref{villa:T8}).  This shows on the
one hand that in spite of the fact that we use 1-dimensional Morse
functions, the contributions coming from the unphysical region
$(-\infty < x < 0 )$ are entirely negligible, and on the other, that
our simple procedure achieves results with a similar accuracy as those
arising from the numerically intensive and sophisticated methods.

The specific transitions which are included for comparisons are:
\begin{enumerate}
\item $A^1 \Sigma^+_u \ - \ X^1 \Sigma^+_g$ in  $^7$Li$_2$  molecule
  (Table \ref{villa:T5}).
\item $B^1 \Pi^+_u \ - \ X^1 \Sigma^+_g$ in  $^7$Li$_2$   molecule
  (Table \ref{villa:T6}).
\item $A^2\Pi _2 \ - \ X^2 \Sigma^+$   in  CN molecule (Table \ref{villa:T12}) .
\item $B^3 \Pi_g \ - \ A^3 \Sigma^+_u$ in  N$_2$ molecule (Table \ref{villa:T8}).
\end{enumerate}
Tables \ref{villa:T5}, \ref{villa:T6}, \ref{villa:T12}, \ref{villa:T8}
contain results of calculations arising from analyses of experimental
data \cite{Kusch:77, Hessel:79, McCallum:70, Walden:72, Zare:65},
simple harmonic oscillator approximations \cite{Drallos:86,
  Nicholls:81}, anharmonic approximations \cite{Rivas:92, Palma:92},
graphical integration \cite{Jarmain:54}, asymptotic expansions
\cite{Chang:70} and by the method of confinement of Morse potentials
in a box \cite{Ley-Koo:95}. We emphasize that we are concerned with an
accurate evaluation of the overlap of two Morse functions and thus
comparisons have to be done with similar calculations, in particular
with the results obtained by Ley-Koo and coauthors \cite{Ley-Koo:95}.

\section{Non-Condon Effects}
Our method can be generalized to include non-Condon effects by
taking into account more terms in the expansion of the electric
dipole operator.  Assuming a power expansion in terms of one of the
Morse coordinates, say ${\xi}_1$ \cite{Bunker:98}
\begin{equation}
\label{TransOp}
{\cal T}({\xi}_1) = \sum_{\eta=0}^{\infty}\, a_\eta ({\xi}_1)^\eta
\, ,
\end{equation}
and following a similar calculation to that leading to (\ref{f12}), we
obtain for the dipole transition

\begin{eqnarray}
  \label{eq:nonC}
&\langle j_2,\nu_2 \vert {\cal T} \vert j_1,\nu_1\rangle &\equiv
\int_{-\infty}^{+\infty}\, \psi_{j_2 \nu_2}^*(x) \, {\cal
T}({\xi}_1(x))\,  \psi_{j_1  \nu_1}(x)\, dx
\\ &=&
\hspace{-40pt}
 2
\sqrt{
{ a \nu_1 ! \nu_2 !  (j_1 - \nu_1) (j_2 -\nu_2)
\over
\Gamma (2 j_1 - \nu_1 + 1) \Gamma (2 j_2 - \nu_2 +1)}
} {y_1}^{j_1-\nu_1} {y_2}^{j_2-\nu_2} \,
\sum^{\nu_1, \nu_2}_{{\ell}_1 , {\ell}_2 = 0 }
{( - 1)^{{\ell}_1 + {\ell}_2}  \over {\ell}_1! {\ell}_2 !}
 \nonumber \\[10pt] && \hspace{-70pt}
\left(
\begin{array}{c}
2j_1-\nu_1 \\ \nu_1 - {\ell}_1
\end{array}
\right)
\left(
\begin{array}{c}
2j_2 - \nu_2 \\ \nu_2 -{\ell}_2
\end{array}
\right) \, {y_1}^{\ell_1} {y_2}^{\ell_2} \,
\times \ \sum_{\eta=0}^{\infty} a_{\eta} \cdot \Bigg\{
 {\cal I}_a ( k_1 + \eta, k_2, y_1,y_2) \Bigg\}
\, ,\nonumber
\end{eqnarray}
with $a=\frac{\beta_2}{\beta_1}$, $k_1=j_1 - \nu_1 + \ell_1$ and $k_2
= j_2 - \nu_2 + \ell_2$. The term $\eta=0$ corresponds to the FC part.
The only difference with (\ref{f12}) is that instead of a single
integral ${\cal I}_a(k_1,k_2,y_1,y_2)$ we have a sum of integrals
${\cal I}_a(k_1+\eta,k_2,y_1,y_2)$, each of them multiplied by the
corresponding coefficient in the expansion (\ref{TransOp}). That means
that we only have to calculate some more integrals and multiply
them by some coefficients which should be chosen in a suitable way.\\

In Figure 1 we have plotted the calculated emission intensities for
the transition ${\tilde C}^1 A' - {\tilde X}^1 A'$ of the $S_2O$
molecule in the $SS$ mode using (a) a zero order approximation (FC),
and (b) including the first correction in (\ref{eq:nonC}) for some
values of the coefficient $a_1$.  The emission intensities were
calculated for the transitions ($0\to \nu$) and ($1 \to \nu$).  We can
see that the effect of the first non-Condon term reduces, as expected,
the transition intensities to the higher levels (large $\nu$) being
more important for the transitions ($1 \to \nu$).  Although we have
not attempted to perform a detailed analysis of this transitions
(since in any case a full polyatomic treatment is necessary), it is
clear that the present approach can be easily adapted to include
non-Condon effects without major changes in the method.

\section{Summary and Conclusions}

In this paper we have calculated Franck-Condon factors by means of a
simple formula (\ref{f12}) (derived from the Wigner function) which
can be adapted in a straightforward manner to include non-Condon
effects when the transition operator is written as a series expansion
in powers of one of the Morse coordinates (\ref{eq:nonC}). The
formula obtained involves well behaved one-dimensional integrations
(\ref{Ia}) which are not tabulated, but for which a symmetry property
(\ref{symmetry}) and a three term recurrence relation
(\ref{recurrence}) have been derived to check the accuracy of their
numerical evaluation. We found that the best strategy to treat the
numerical integrations is by means of a scaling method in which the
integral is transformed in such a way that the integrand acquires a
unit value at the maximum shifted at $x=1$.  This approach is
particularly useful when the transitions involve large bonding
energies (or large values of the parameters $j_1, j_2$).\\

Since the present approach is oriented to possible applications to the
calculation of multidimensional Franck-Condon factors, we have
confined our study to the calculation of one-dimensional Morse
function overlaps by using a simple program written in {\it
  Mathematica} (see appendix B for a listing of the
{\it Mathematica} program used to calculate the FC factors). \\

As far as we only require transitions between the
lowest-lying states in each Morse potential (up to $\nu_1 \approx
\nu_2 \approx 10$) our results have a similar accuracy as those
obtained through more sofisticated methods.  Although in this work we
have included only a few examples, we were able to reproduce all
results in \cite{Ley-Koo:95} with high precision. Only for cases
involving long summations we observe some isolated small deviations.
We emphasize that in the above transitions we are dealing with rather
large values of the bonding energies in the Morse potentials (i.e.
large summations in eq. (\ref{f12})), which suggests that our
procedure can be reliably applied to a wide variety of physical
examples. On the basis of these results, it is our impression that the
inaccuracies arising from the alternating sums in (\ref{f12}) have
been overestimated.  Such sums can be calculated by {\it Mathematica}
with no significant loss of precision for a wide range of problems of
physical interest.  However, for transitions involving higher-states a
more elaborate programming of our formula is required in order to avoid
the loss of accuracy. This is beyond the objective of the present
work and will be studied elsewhere. \\

In conclusion, Franck-Condon factors for the Morse potential and
corrections including non-Condon effects can be calculated with high
accuracy by means of a very simple program suggesting that this
approach is very convenient, since it can be implemented in
any personal computer.\\

\newpage
\def\theequation{\thesection.\arabic{equation}}
\appendix
\setcounter{equation}{0}

\section{Derivation of Franck-Condon overlaps for Morse Potential eigenfunctions}

Let's consider the Wigner function of two different
Morse wavefunctions  $\psi^{\beta_1}_{j_1\nu_1} (x)$ and
$\psi^{\beta_2}_{j_2\nu_2} (x)$ (\ref{Wigner}),

\begin{eqnarray}
   W( \psi_{j_1,\nu_1}^{\beta_1}, \psi_{j_2,\nu_2}^{\beta_2} \vert x,p )
&=&
\frac{{\cal N}_{j_1,\nu_1}^{\beta_1}{\cal N}_{j_2,\nu_2}^{\beta_2}
  }{2\pi \hbar} \int_{-\infty}^{\infty} dr\,
{\rm exp}\bigg\{
-\frac{1}{2} (2 j_1 + 1) e^{\displaystyle -\beta_1(x-\frac{1}{2}r)}
\bigg\} \nonumber \\
&& \bigg[
(2 j_1 + 1) e^{\displaystyle -\beta_1(x-\frac{1}{2}r)}
\bigg]^{j_1-\nu_1} \,
L_{\nu_1}^{2j_1-2\nu_1}\left[ (2 j_1 + 1)
e^{-\beta_1(x-\frac{1}{2}r)} \right]
\nonumber \\
&& e^{-i\, p r /\hbar} \
{\rm exp}\bigg\{
-\frac{1}{2} (2 j_2 + 1) e^{\displaystyle -\beta_2(x+\frac{1}{2}r)}
\bigg\}
\nonumber \\
&& \bigg[
(2 j_2 + 1) e^{\displaystyle -\beta_2(x+\frac{1}{2}r)}
\bigg]^{j_2-\nu_2} \,
L_{\nu_2}^{2j_2-2\nu_2}\left[ (2 j_2 + 1)
e^{-\beta_2(x+\frac{1}{2}r)} \right] \, , \nonumber
\end{eqnarray}
defining
\[
y_1 = (2j_1+1) e^{-\beta_1 x}, \quad y_2 =  (2j_2+1) e^{-\beta_2 x},
\]
and making the change of variables
\[
u= e^{\beta_1\, r /2},
\quad
e^{\beta_2\, r /2} = u^{{\beta_2}/{\beta_1}},
\quad
du = \frac{\beta_1}{2} u \, dr\,  ,
\]
we have
\begin{eqnarray}
  W( \psi_{j_1,\nu_1}^{\beta_1}, \psi_{j_2,\nu_2}^{\beta_2} \vert x,p )
&=& \frac{{\cal N}_{j_1,\nu_1}^{\beta_1}{\cal N}_{j_2,\nu_2}^{\beta_2}
  }{2\pi \hbar}
y_1^{(j_1-\nu_1)} y_2^{(j_2-\nu_2)}\nonumber \\
&& \int_{-\infty}^{\infty} du\, {\rm exp} \big\{  -\frac{1}{2} y_1 u -
\frac{1}{2} y_2 u^{-\beta_2/\beta_1} \big\}\,
u^{(j_1-\nu_1-\frac{\beta_2}{\beta_1}(j_2-\nu_2)- \frac{2 i p
    }{\beta_1 \hbar} ) } \nonumber \\
&& L_{\nu_1}^{2j_1 -2\nu_1} (y_1 u) \, L_{\nu_2}^{2j_2 -2\nu_2} (y_2
u^{-\frac{\beta_2}{\beta_1}}) \, ,
\nonumber
\end{eqnarray}
then, expanding the associated Laguerre polynomials in their finite
series (\ref{Laguerre}), we arrive finally to
\begin{eqnarray}
  W( \psi_{j_1,\nu_1}^{\beta_1}, \psi_{j_2,\nu_2}^{\beta_2} \vert x,p )
&=& \frac{{\cal N}_{j_1,\nu_1}^{\beta_1}{\cal N}_{j_2,\nu_2}^{\beta_2}
  }{\pi \beta_1\hbar}
y_1^{(j_1-\nu_1)} y_2^{(j_2-\nu_2)}\nonumber \\
&&  \sum^{\nu_1, \nu_2}_{{\ell}_1 , {\ell}_2 = 0 }
{( - 1)^{{\ell}_1 + {\ell}_2}  \over {\ell}_1! {\ell}_2 !}
\left(
\begin{array}{c}
2j_1-\nu_1 \\ \nu_1 - {\ell}_1
\end{array}
\right)
\left(
\begin{array}{c}
2j_2 - \nu_2 \\ \nu_2 -{\ell}_2
\end{array}
\right) \, {y_1}^{\ell_1} {y_2}^{\ell_2} \,
\nonumber \\
&& \hspace{-40pt}\int_{0}^{\infty} du\, {\rm exp} \big\{  -\frac{1}{2} y_1 u -
\frac{1}{2} y_2 u^{-\beta_2/\beta_1} \big\}\,
u^{(\displaystyle j_1-\nu_1-\frac{\beta_2}{\beta_1}(j_2-\nu_2)- \frac{2 i p
    }{\beta_1 \hbar} +{\ell}_1 -\frac{\beta_2}{\beta_1}{\ell}_2 -1) }
\, , \nonumber
\end{eqnarray}
from which we obtain the Franck-Condon factors (\ref{f12}).

\setcounter{equation}{0}
\section{The {\it Mathematica} program for the calculation of
the Franck-Condon factors with wave functions of the Morse
potential.} 

\small
\begin{verbatim}
FCFact[j1_,nu1_,beta1_,R1_,j2_,nu2_,beta2_,R2_,flag_]:=
  Module[
    {y1,y2,a},
    (* --------------------------------------------------------------------- *)
    (* This program computes the Franck-Condon factor of two Morse functions *)
    (* Usage:                                                                *)
    (* FCFact[j1,nu1,beta1,R1,j2,nu2,beta2,R2,flag]                          *)
    (* flag=0 computes the FC factor                                         *)
    (* flag=1 computes the square of the FC factor                           *)
    (* --------------------------------------------------------------------- *)
    (* Definition of Local Variables *)
    Acc=12;
    AccN=100;
    a= SetAccuracy[beta2/beta1,AccN];
    y1=SetAccuracy[(2*j1+1)*Exp[-beta1*(R2-R1)/2],AccN];
    y2=SetAccuracy[(2*j2+1)*Exp[-beta2*(R1-R2)/2],AccN];
    Fcsum=0;
    fac0a=SetAccuracy[2*Sqrt[a*FullSimplify[nu1!*(j1-nu1)*nu2!*(j2-nu2)/
                          (Gamma[2*j1-nu1+1]*Gamma[2*j2-nu2+1])]],AccN];
    fac0b=SetAccuracy[y1^(j1-nu1)*y2^(j2-nu2),AccN];
    fac0=fac0a*fac0b;
     Do[ (* Do loops for summations *)
      fac1a=SetAccuracy[(2*j1+1)^(l1)*(2*j2+1)^(l2),AccN];
      fac1b=SetAccuracy[Exp[-beta1*(R2-R1)*(l1)/2]*Exp[-beta2*(R1-R2)*(l2)/2]
            ,AccN];
      fac1=SetAccuracy[fac1a*fac1b,AccN];
      fac2=SetAccuracy[(-1)^(l1+l2)/(l1!*l2!),AccN];
      fac3=SetAccuracy[FullSimplify[Binomial[2*j1-nu1,nu1-l1]*
                                    Binomial[2*j2-nu2,nu2-l2]],AccN];
      Fcaux=SetAccuracy[fac0 *fac1*fac2*fac3,AccN];
      (* =========================================================== *)
      (* Integration Section. Integral Ia. Calculation with scaling  *)
      (* =========================================================== *)
      Lambda=SetAccuracy[j1-nu1+l1 + a*(j2-nu2+l2 ),AccN];
      (* Integrand *)
      integr=x^(Lambda-1)*Exp[-(y1*x+y2*x^a)/2];
      finteg[xx_]:=xx^(Lambda-1)*Exp[-(y1*xx+y2*xx^a)/2];
      (* Looking for position of maximum *)
      aux=FindMinimum[-integr,{x,1,0,1000},PrecisionGoal->10];
      peak=aux[[2]][[1]][[2]];
        xatmax =SetAccuracy[peak,AccN]; (* Position of maximum *)
        (* Integrand value at maximum *)
        fatmax=SetAccuracy[finteg[peak],AccN];
        finteg2=SetAccuracy[(1/fatmax)*(X*xatmax)^(Lambda-1)*
                Exp[-(y1*(X*xatmax)+y2*(X*xatmax)^a)/2],AccN];
        res=SetAccuracy[fatmax*xatmax*NIntegrate[finteg2,{X,0,Infinity},
            PrecisionGoal->Acc,AccuracyGoal->Acc,
            Method->GaussKronrod],AccN];
      (* Adding terms *)
      Fcsum=Fcsum+SetAccuracy[Fcaux*res,AccN]
      ,{l1,0,nu1},{l2,0,nu2}];
    (* Choosing Output *)
    If[flag==0, FCfactor=Fcsum];
    If[flag==1, FCfactor=Fcsum^2];
    N[FCfactor,15]
    ]
\end{verbatim}

\newpage
\pagestyle{empty}

\newpage

\pagestyle{empty}
%
\begin{center}
  {\bf \Large FIGURES }
\end{center}

\newpage

\begin{figure}[htb]
    \label{fig:S2Oss}
  \begin{center}
\[
\begin{array}{c}
\psfig{figure=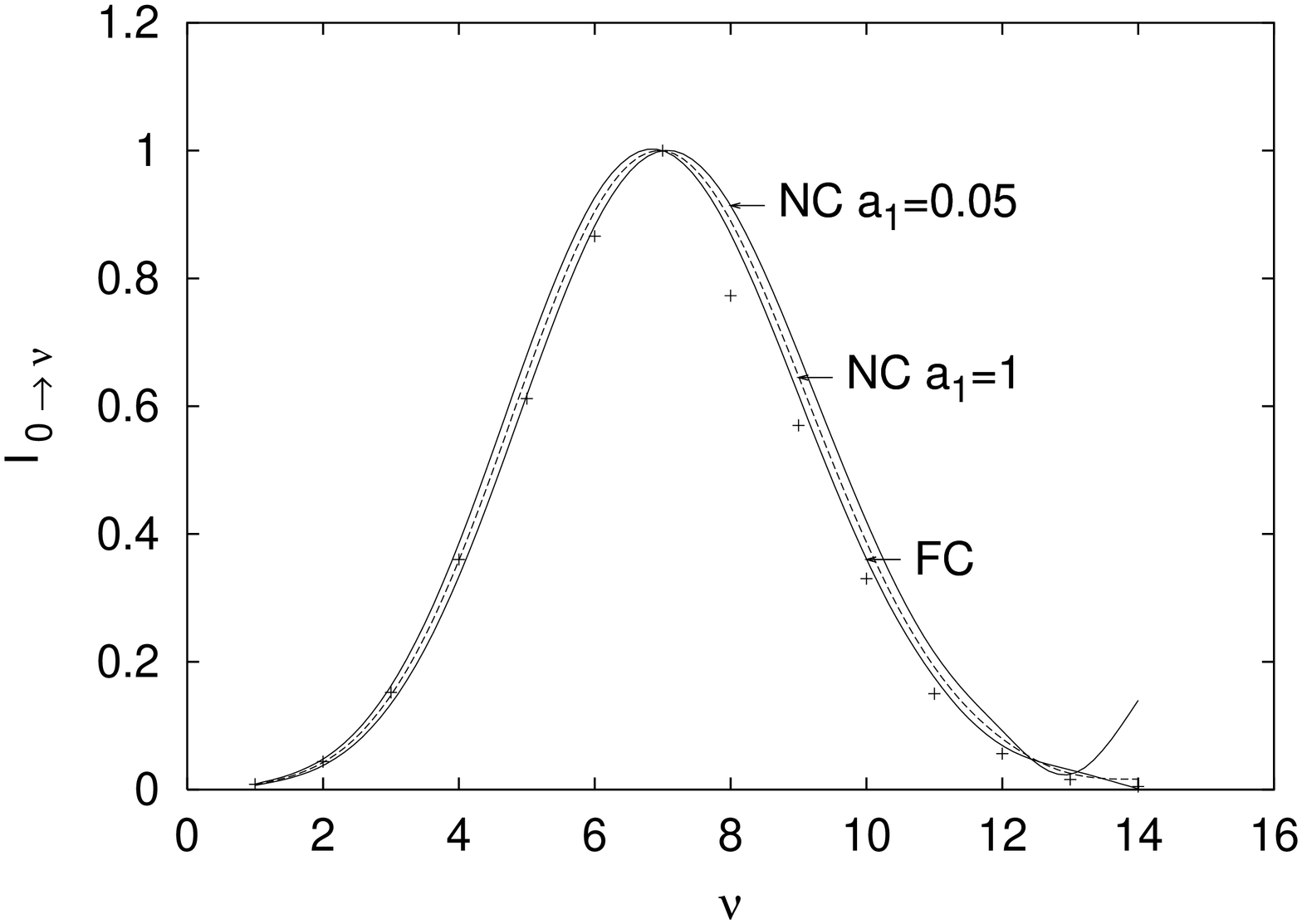,height=4.5in,angle=-90} \\[5pt]
(a) \\[10pt]
\psfig{figure=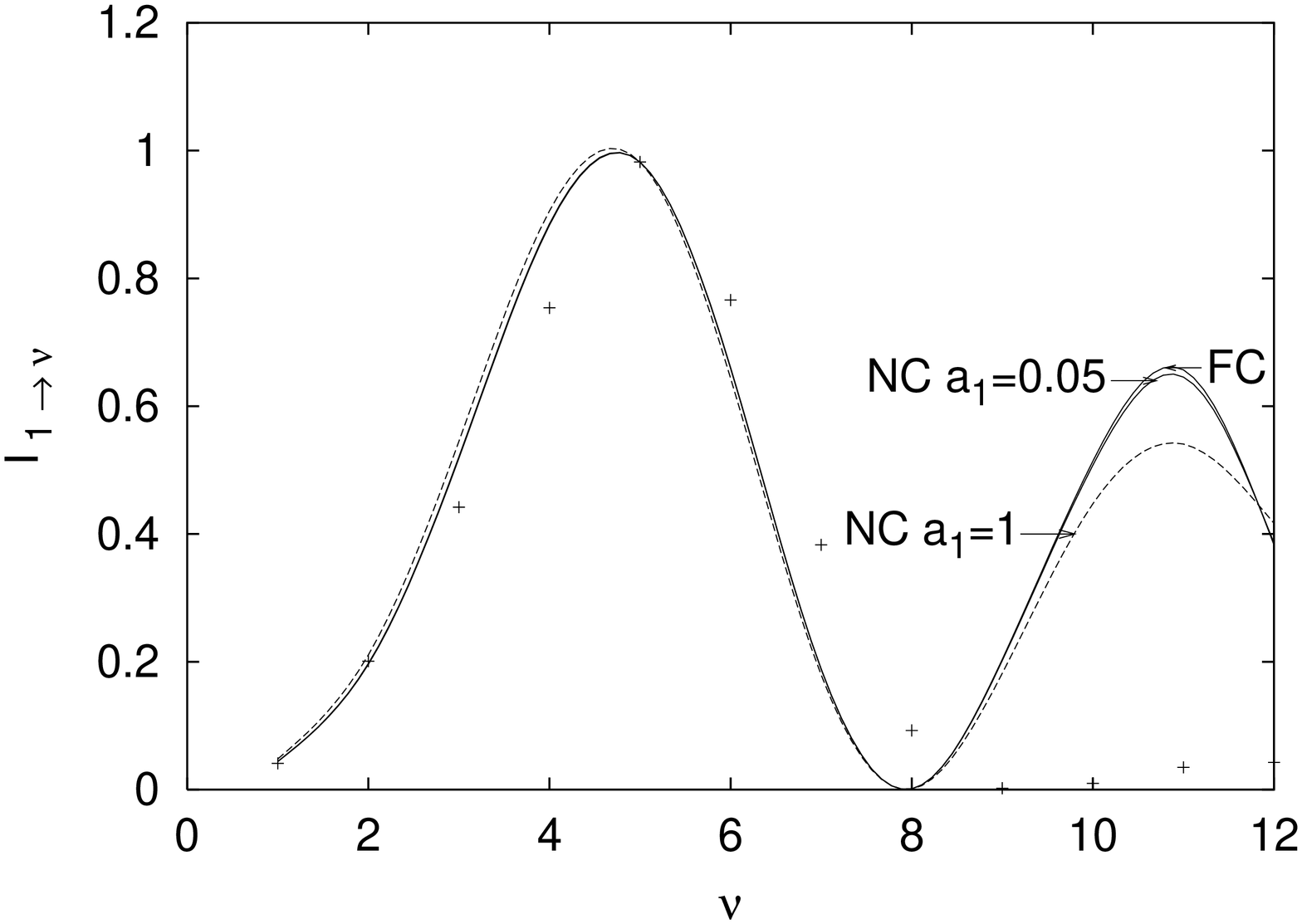,height=4.5in,angle=-90} \\[5pt]
(b) \\
\end{array}
\]
    \caption{Emission intensities in the $S_2O$ ${\tilde C}^1 A' -
      {\tilde X}^1 A'$ transitions (a) $I_{0\to\nu}$ and (b)
      $I_{1\to\nu}$. The calculations in the Condon approximation
      using Eq.(\ref{f12}) (FC) as well as the calculations with
      inclusion of the first non-Condon correction as given by
      Eq.(\ref{eq:nonC}) (NC) for $a_1=0.01$ and $a_1=1$ are displayed.
      Data from experimental analysis indicated by crosses is from
      ref.\cite{Muller:99}.  The parameters for this calculations are
      $j_1=128,\beta_1=1.6,j_2=79,\beta_2=1.8$ and
      $\Delta=R_2-R_1=0.26$ }
  \end{center}
\end{figure}

\clearpage

%
\begin{center}
  {\bf \Large TABLES }
\end{center}

\newpage

\begin{table}[htbp]
  \begin{center}
  {\small
    \begin{tabular}{|c|c|c|c|c|c|c|c|c|} \hline
$\nu_{\widetilde X} \diagdown \nu_{\widetilde C}$ & 0 & 1 & 2 & 3 & 4 & 5 & 6 \\ \hline
0& 0.002750 &  -0.007793 &  0.016082 &  -0.027885  &  0.043045  & -0.061043 &  0.081101 \\
 & 0.002750 &  -0.007793 &  0.016082 &  -0.027885  &  0.043045  & -0.061043 &  0.081101 \\
 & 0.003062 &  -0.008764 &  0.018342 &  -0.032357  &  0.050950  & -0.073850 &  0.100423 \\ \hline
1& 0.011185 &  -0.029081 &  0.054964 &  -0.087075  & 0.122463   & -0.157692 &  0.189462 \\
 & 0.011185 &  -0.029081 &  0.054964 &  -0.087075  & 0.122463   & -0.157692 &  0.189462 \\
 & 0.011878 &  -0.031187 &  0.059745 &  -0.096203  & 0.137787   & -0.180866 &  0.221525 \\ \hline
2& 0.031189 &  -0.073353 &  0.124795 &  -0.176850  &  0.220705  & -0.249527 &  0.259511 \\
 & 0.031189 &  -0.073353 &  0.124795 &  -0.176850  &  0.220705  & -0.249527 &  0.259511 \\
 & 0.031718 &  -0.075369 &  0.130022 &  -0.187312  & 0.237944   & -0.273787 &  0.289206 \\ \hline
3& 0.068791 &  -0.143597 &  0.214500 &  -0.262849  &  0.277331  & -0.255821 &  0.203940 \\
 & 0.068791 &  -0.143597 &  0.214500 &  -0.262849  &  0.277331  & -0.255821 &  0.203941 \\
 & 0.067240 &  -0.142034 &  0.215536 &  -0.269007  &  0.289363  & -0.271747 &  0.219257 \\ \hline
4& 0.127150 &  -0.229507 &  0.289495 &  -0.287861  &  0.228221  & -0.130373 &  0.020252 \\
 & 0.127150 &  -0.229507 &  0.289495 &  -0.287861  &  0.228221  & -0.130372 &  0.020249 \\
 & 0.119888 &  -0.219637 &  0.282588 &  -0.287765  & 0.234410   & -0.138111 &  0.023018 \\ \hline
5& 0.203169 &  -0.305290 &  0.303169 &  -0.208109  &  0.065636  &  0.072441 &  -0.168141 \\
 & 0.203169 &  -0.305290 &  0.303169 &  -0.208109  &  0.065636  &  0.072439 & -0.168110 \\
 & 0.185445 &  -0.284285 &  0.290441 &  -0.207870  &  0.073229  &  0.066095 & -0.170366 \\ \hline
6& 0.286075 &  -0.337012 &  0.224040 &  -0.035244  &  -0.130845 &  0.213879 &  -0.202840 \\
 & 0.286075 &  -0.337012 &  0.224040 &  -0.035245  &  -0.130843 &  0.213905 & -0.203158 \\
 & 0.253653 &  -0.307590 &  0.215240 &  -0.045192  &  -0.116517 &  0.208866 & -0.211595 \\ \hline
7& 0.359518 &  -0.297912 &  0.062331  &  0.153092   &  -0.235782 &  0.181032 &  -0.053809 \\
 & 0.359518 &  -0.297912 &  0.062329  &   0.153106  & -0.235815  &  0.180778 &  -0.049421  \\
 & 0.310715 &  -0.269700 &  0.069937  &   0.129446  & -0.222370  &  0.188315 &  -0.070168  \\ \hline
    \end{tabular}
    \caption{Franck-Condon factors (\ref{f12}) for $\nu_{\widetilde X}
      = 0\ldots 7$ and $\nu_{\widetilde C} = 0\ldots 6$ for the $S-S$
      stretching degree of freedom in the $S_2O$ molecule. Parameters
      for overlaps are $j_1=128,\beta_1=1.6,j_2=79,\beta_2=1.8$ and
      $\Delta=R_2-R_1=0.283$.  Successive entries correspond to the
      calculations using formula (\ref{f12}) with a scaled integation
      (first line), using the Configuration Localized States method of
      ref.\cite{Carvajal:99, curro:99} and using the algebraic formula of
      ref.\cite{Iach:JCP98}.}
    \label{tab:curroT3}
}
  \end{center}
\end{table}

\begin{table}[h]
\begin{center}
\begin{tabular}{|c|c|c|c|c|c|} \hline
$\nu_1 \diagdown \nu_2$ & 0  & 1  & 2 & 3  & 4  \\
\hline \hline
0 & 0.42687 & -0.34228 & 0.27839 & -0.21679 & 0.14889 \\
  & 0.38163 & -0.52971 & 0.51990 & -0.41663& 0.28915  \\
  & 0.38163 & -0.45147 & 0.49396 & -0.52002 & 0.50695 \\  \hline
1 & 0.79048 & -0.12775 & 0.04693 & -0.00110 & - 0.01166 \\
  & 0.52971 & -0.35362 & -0.02750 & 0.32220 & -0.43193 \\
  & 0.62781 & -0.28358 & 0.05443 & 0.16330 & - 0.38378 \\  \hline
2 & 0.41678 & 0.74799 & -0.13014 & 0.13408 & -0.08036 \\
  & 0.51990 & 0.02750 & -0.38061 & 0.28255 & 0.03174 \\
  & 0.23717 & 0.51298 & -0.40591 & 0.24204 & 0.01297 \\  \hline
3 &-0.12667 & 0.44613 & 0.77692 & 0.04596 & 0.09256 \\
  & 0.41663 & 0.32220 & -0.8255 & -0.15417 & 0.35170 \\
  & -0.09261 & 0.32338 & 0.41678 & -0.40496 & 0.29047 \\ \hline
4 & 0.02807 & -0.24277 & 0.21095 & 0.81696 & 0.36737 \\
  & 0.28915 & 0.43193 & 0.03174 & -0.35170 & 0.08991 \\
  & 0.05862 & -0.19434 & 0.36082 & 0.38247 & -0.39483 \\ \hline
\end{tabular}
\end{center}
\caption{Franck-Condon overlap integrals
$f_{1,2}$ between states in
identical Morse  potentials displaced one with respect to the other.
The Morse parameters are $j_1 = j_2 =5$, $\beta_1 = \beta_2 =
0.90$ $\mbox{\AA}^{-1}$,  $R_1 =2.67 \mbox{\AA}$ and  $R_2 = 3.60
\mbox{\AA}$. The first line correspond to the results of formula
(\ref{f12}) using numerical integrations (\ref{Ia}). Identical
results are obtained with the exact formula (\ref{f12a1}), and are
also identical to the results of \cite{Carvajal:99}.
Second line corresponds to
the calculations using  harmonic oscillator wave
functions with oscillator length $a_0 = [(j + {1\over 2})^{1/2}\beta]^{-1}$,
 and third line shows the calculations with modified harmonic oscillator
wave functions as given in  Ref.\cite{Iach:JCP98}. }
\label{carvaj:TI}
\end{table}


\clearpage

\begin{table}[h]
\begin{center}
\begin{tabular}{|c|c|c|c|c|c|} \hline
$\nu_1 \diagdown \nu_2$ & 0   & 1 & 2  &  3   &  4  \\
\hline
0 & 0.52851 & -0.48137 & 0.37794 & -0.27795 & 0.18265 \\
  & 0.53110 & -0.68046 & 0.47203 & -0.13547 & -0.07045 \\
  & 0.53108 & -0.61175 & 0.61258 & -0.55162 & 0.40091 \\
\hline
1 & 0.69643 & -0.11240 & -0.11346 & 0.16657 & -0.13819 \\
  & 0.45363 & -0.09097 & -0.48511 & 0.63898 & -0.31027 \\
  & 0.47415 & -0.11648 & -0.01808 & 0.41928 & -0.56771 \\
\hline
2 & 0.47491 & 0.57459 & -0.30630 & 0.12814 & -0.04358 \\
  & 0.41843 & 0.05608 & -0.24858 & -0.19935 & 0.62758 \\
  & 0.29134 & 0.25698 & -0.25966 & 0.08329 & 0.17547 \\
\hline
3 & 0.09588 & 0.64689 & 0.47316 & -0.26552 & 0.15657 \\
  & 0.34881 & 0.22208 & -0.23267 & -0.12709 & -0.00045 \\
  & 0.06602 & 0.38994 & -0.02755 & -0.11715 & 0.08620 \\
\hline
4 & -0.02603 & 0.04474 & 0.69961 & 0.57982 & -0.08986 \\
  & 0.28834 & 0.27451 & -0.03721 & -0.30668 & 0.09155 \\
  & 0.00203 & 0.13429 & 0.38065 & -0.22119 & 0.09124 \\
\hline
\end{tabular}
\end{center}
\caption{Franck-Condon  overlap integrals $f_{1,2}$ between states in
different Morse potentials. Parameters in  potentials are
$j_1 = 5$, $R_1 = 2.67 \mbox{\AA}$,  $\beta_1 = 0.90
\mbox{\AA}^{-1}$ and $j_2 = 5$, $R_2 = 3.60 \mbox{\AA}$,
$\beta_2 =0.60 \mbox{\AA}^{-1}$.  Successive  entries
correspond to results of formula (\ref{f12}) (which agree
exactly with  the  calculations of ref.\cite{Carvajal:99}
making the integration of the Morse wave functions),
the calculation using harmonic-oscillator wave functions
with oscillator length $a_0=[(j+\frac{1}{2})^{1/2} \beta]^{-1}$,
and the calculation with modified harmonic oscillator
wave functions as given in ref.\cite{Iach:JCP98}.}
\label{carvaj:TII}
\end{table}

\clearpage

\begin{table}[h]
\begin{center}
{\small
\begin{tabular}{|c|l|l|l|l|l|l|l|} \hline
$\nu_1 \diagdown \nu_2$ & 0    &  1  &  2   &  3   &  4  & 5 & 6   \\
\hline
0 & 5.200 -- 2 & 1.340 -- 1 & 1.870 -- 1 & 1.900 -- 1 & 1.570 -- 1 & 1.130-- 1 & 7.300 -- 2 \\
  & 5.300 -- 2 & 1.310 -- 1 & 1.820 -- 1 & 1.880 -- 1 & 1.580 -- 1 &-- & -- \\
  & 5.400 -- 2 & 1.320 -- 1 & 1.830 -- 1 & 1.880 -- 1 & 1.580 -- 1 &-- & -- \\
  & 5.603 -- 2 & 1.396 -- 1 & 1.916 -- 1 & 1.919 -- 1 & 1.566 -- 1 &1.105 -- 1 & 6.990 -- 2 \\
  & .0560317  & .1395728  & .1916222  & .1918638  & .1565977  & .1105150 & .0699024 \\
\hline
1 & 1.760 -- 1 & 1.970 -- 1 & 7.900 -- 2 & 3.000 -- 3 & 1.800 -- 2 &6.800 -- 2 & 1.020 -- 1 \\
  & 1.800 -- 1 & 1.910 -- 1 & 7.800 -- 2 & 4.000 -- 3 & 1.500 -- 2 &-- & -- \\
  & 1.820 -- 1 & 1.910 -- 1 & 7.700 -- 2 & 4.000 -- 3 & 1.600 -- 2 &--  & -- \\
  & 1.927 -- 1 & 1.983 -- 1 & 7.046 -- 2 & 1.277 -- 3 & 2.431 -- 2 &7.603 -- 2 & 1.060 -- 1 \\
  & .1927494  &  .1983307  &  .0704506  &  .0012764 &  .0243160  &  .0760309  &  .1060282  \\
\hline
2 & 2.700 -- 1 & 5.800 -- 2 & 1.500 -- 2 & 9.800 -- 2 & 9.000 -- 2 &2.700 -- 2 & -- \\
  & 2.780 -- 1 & 5.400 -- 2 & 1.300 -- 2 & 9.000 -- 2 & 8.900 -- 2 &-- & -- \\
  & 2.780 -- 2 & 5.200 -- 2 & 1.500 -- 2 & 9.200 -- 2 & 9.000 -- 2 &-- & -- \\
  & 2.926 -- 1 & 4.405 -- 2 & 2.609 -- 2 & 1.062 -- 1 & 8.071 -- 2 &1.721 -- 2 & 1.653 -- 3 \\
  & .2926194  &  .0440430  &  .0261028  & .1062452  &  .0807091  &  .0172068  &  .0016530 \\
\hline
3 & 2.500 -- 1 & 9.000 -- 3 & 1.270 -- 1 & 4.500 -- 2 & 4.000 -- 3 &6.000 -- 2 & -- \\
  & 2.540 -- 1 & 1.200 -- 2 & 1.200 -- 1 & 4.600 -- 2 & 2.000 -- 3 &-- & -- \\
  & 2.530 -- 1 & 1.300 -- 2 & 1.210 -- 1 & 4.400 -- 2 & 2.000 -- 3 &-- & -- \\
  & 2.555 -- 1 & 2.655 -- 2 & 1.349 -- 1 & 2.666 -- 2  & 1.349 -- 2 &7.235 -- 2 & 6.873 -- 2 \\
  &  2.551 -- 1 & 2.632 -- 2 & 1.335 -- 1 & 2.931 -- 2 & 2.025 -- 3 &   7.142 -- 2 & 6.810 -- 2 \\
& .2554832 &  .0265683 &  .1349060 &  .0266544 &  .0134925 &  .0723511 &  .0687292 \\
\hline
4 & 1.560 -- 1 & 1.340 -- 1 & 5.600 -- 2 & 2.500 -- 2 & 9.200 -- 2 &   3.000 -- 2 & 2.000 -- 3 \\
  &  1.530 -- 1 & 1.450 -- 1 & 5.100 -- 2 & 2.000 -- 2 & 8.400 -- 2 & --& -- \\
  &  1.510 -- 1 & 1.480 -- 1 & 4.900 -- 2 & 2.100 -- 2 & 8.500 -- 2 & --& -- \\
  & 1.403 -- 1 & 1.851 -- 1 & 2.840 -- 2 & 5.294 -- 2 & 8.777 -- 2 &1.151 -- 2 & 1.385 -- 2 \\
  & .1403039  & .1851266  &  .0283826  &  .0529617  &  .0877665  &  .0115032  &  .0138506  \\
\hline
5 & 6.800 -- 2 & 2.110 -- 1 & 8.000 -- 3 & 1.120 -- 1 & 9.000 -- 3 &3.900 -- 2 & 7.100 -- 2 \\
& -- & -- & -- & --  & -- & -- & -- \\
& -- & -- & -- & --  & -- & -- & -- \\
& 4.983 -- 2 & 2.285 -- 1 & 4.430 -- 2 & 1.025 -- 1 & 2.335 -- 4 &6.681 -- 2 & 5.516 -- 2 \\
& .0498214  & .2285020  &  .0443334  &  .1024608  &  .0002352  &  .0668249  &  .0551753  \\
\hline
6 & 2.100 -- 2 & 1.590 -- 1 & 1.290 -- 1 & 2.500 -- 2 & 6.100 -- 2 &6.200 -- 2 & 2.000 -- 3 \\
  & -- & -- & -- & -- & -- & -- & -- \\
  & -- & -- & -- & -- & -- & -- & -- \\
  & 1.132 -- 2 & 1.302 -- 1 & 2.046 -- 1 & 1.996 -- 4 & 1.011 -- 1 &2.514 -- 2 & 1.754 -- 2 \\
  & .0113133   &  .1301468   &  .2046446   &  .0001973   &  .1011479   &  .0251204   &  .0175930   \\
\hline
\end{tabular}
}
\end{center}
\caption{Franck-Condon (squared) overlap between states
involved in the transition $A^1 \Sigma^+_u \ - \ X^1 \Sigma^+_g$
of $^7$Li$_2$ molecule. The parameters of this transition are:
$j_1 = 69.494$, $\beta_1 = 0.326 \ a_0^{-1}$, $R_{1} = 5.876 \ a_0$.
$j_2 = 48.106$, $\beta_2 = 0.459 \ a_0^{-1}$, $R_{2} = 5.0535 \ a_0$.
Successive entries  correspond to:
Kusch\cite{Kusch:77}, Drallos\cite{Drallos:86},
Rivas-Silva\cite{Rivas:92},
Ley-Koo\cite{Ley-Koo:95} and
our results using formula (\ref{f12}) with the scaled integral
(\ref{Ia}).}
\label{villa:T5}
\end{table}

\clearpage

\begin{table}[h]
\begin{center}
{\small
\begin{tabular}{|c|l|l|l|l|l|l|l|} \hline
$\nu_1 \diagdown \nu_2$ & 0    &  1  &  2   &  3
   &  4  & 5 & 6   \\
\hline
0 & 3.188 -- 1 & 3.340 -- 1 & 2.088 -- 1 & 9.180 -- 2 & 3.580 -- 2 &1.260 -- 2 & 4.200 -- 3 \\
  & 3.267 -- 1 & 4.104 -- 1 & 2.065 -- 1 & 5.070 -- 2 & 5.500 -- 3 &-- & -- \\
  & 3.222 -- 1 & 4.090 -- 1 & 2.094 -- 1 & 5.300 -- 2 & -- & -- & --\\
  & 3.382 -- 1 & 3.431 -- 1 & 1.963 -- 1 & 8.285 -- 2 & 2.850 -- 2 &8.369 -- 3 & 2.138 -- 3 \\
&.33817247 & .34308979 & .19630112 & .08284818 & .02850117 & .00836875 & .00213843 \\
\hline
1 & 3.827 -- 1 & 7.700 -- 3 & 9.420 -- 2 & 1.884 -- 1 & 1.585 -- 1 &9.290 -- 2 & 4.460 -- 2 \\
  & 3.149 -- 1 & 3.900 -- 3 & 2.042 -- 1 & 3.127 -- 1 & 1.395 -- 1 &-- & -- \\
  & 3.154 -- 1 & 4.900 -- 3 & 1.972 -- 1 & 3.117 -- 1 & -- & -- & --\\
  & 3.941 -- 1 & 1.801 -- 3 & 1.255 -- 1 & 2.075 -- 1 & 1.522 -- 1 &7.607 -- 2 & 2.975 -- 2 \\
&.39406016 & .00180094 & .12553500 & .20752753 & .15219139 & .07606690 & .02974802 \\
\hline
2 & 2.103 -- 1 & 1.511 -- 1 & 1.345 -- 1 & 1.000 -- 4 & 7.110 -- 2 &1.374 -- 1 & 1.263 -- 1 \\
  & 1.961 -- 1 & 8.440 -- 2 & 1.110 -- 1 & 3.290 -- 2 & 2.826 -- 1 &-- & -- \\
  & 1.978 -- 1 & 8.210 -- 2 & 1.143 -- 1 & 2.890 -- 2 & -- & -- & --\\
  & 2.011 -- 1 & 1.966 -- 1 & 1.043 -- 1 & 7.137 -- 3 & 1.181 -- 1 &1.612 -- 1 & 1.165 -- 1 \\
&.20107856 & .19664231 & .10427085 & .00713746 & .11809640 & .16124354 & .11651784 \\
\hline
3 & 6.980 -- 2 & 2.711 -- 1 & 6.300 -- 3 & 1.508 -- 1 & 5.500 -- 2 &2.400 -- 3 & 6.310 -- 2 \\
  & 9.690 -- 2 & 1.692 -- 1 & 5.000 -- 5 & 1.391 -- 1 & 2.800 -- 3 &-- & -- \\
  & 9.820 -- 2 & 1.688 -- 1 & 1.000 -- 4 & 1.394 -- 1 & -- & -- & --\\
  & 5.656 -- 2 & 2.865 -- 1 & 3.732 -- 2 & 1.564 -- 1 & 1.533 -- 2 &3.566 -- 2 & 1.220 -- 1 \\
&.05655825 & .28654540 & .03732449 & .15638530 & .01533330 & .03566019 & .12201313 \\
4 & 1.560 -- 2 & 1.657 -- 1 & 1.834 -- 1 & 3.030 -- 2 & 6.610 -- 2 &1.082 -- 1 & 2.040 -- 2 \\
  & 4.130 -- 2 & 1.516 -- 1 & 6.220 -- 2 & 4.230 -- 2 & 8.000 -- 2 &-- & -- \\
  & -- & -- & -- & -- & -- & -- & -- \\
  & 9.241 -- 3 & 1.370 -- 1 & 2.568 -- 1 & 5.520 -- 4 & 1.241 -- 1 &6.346 -- 2 & 1.335 -- 3 \\
&.00924096 & .13696229 & .25680238 & .00055202 & .12409529 & .06346440 & .00133720 \\
\hline
5 & 2.500 -- 3 & 5.600 -- 2 & 2.243 -- 1 & 6.280 -- 2 & 1.028 -- 1 &3.900 -- 3 & 8.950 -- 2 \\
  & -- & -- & -- & -- & -- & -- & -- \\
  & -- & -- & -- & -- & -- & -- & -- \\
  & 8.503 -- 4 & 3.121 -- 2 & 2.068 -- 1 & 1.765 -- 1 & 3.284 -- 2 &6.442 -- 2 & 9.497 -- 2 \\
&.00085027 & .03120572 & .20683663 & .17650291 & .03283566 & .06440957 & .09479757 \\
\hline
\end{tabular}
}
\end{center}
\caption{Franck-Condon (squared) overlap between states
involved in the transition $B^1\Pi^+_u \ - \ X^1 \Sigma^+_g$
of $^7$Li$_2$ molecule. The parameters of this transition are:
$j_1=62.05$, $\beta_1 = 0.354 \ a_0^{-1}$, $R_{1} = 5.550 \ a_0$.
$j_2=48.104$, $\beta_2 = 0.459 \ a_0^{-1}$, $R_{2} = 5.0535 \ a_0$.
Successive entries  correspond to:
Hessel\cite{Hessel:79}, Drallos\cite{Drallos:86},
Palma\cite{Rivas:92}, Ley-Koo\cite{Ley-Koo:95}  and our results
using formula (\ref{f12}) with the scaled integral
(\ref{Ia}).}
\label{villa:T6}
\end{table}

\clearpage

\begin{table}[h]
\begin{center}
{\small
\begin{tabular}{|c|l|l|l|l|} \hline
$\nu_1 \diagdown \nu_2$ & 0   &  1   &  2    &  3      \\
\hline
0 & .4960     & .3370     & .1220      & .0280 \\
  & .4990     & .3200     & .1260      & .0399 \\
  & .4940     & .3350     & .1260      & .0349 \\
  & .4970     & .3236     & .1279      & .0389 \\
  & .5047     & .3295     & .1222      & .0339 \\
  & .50469840 & .32944876 & .12216458  & .03391230
\\ \hline
1 & .3470     & .0450     & .2940     & .2150 \\
  & .3710     & .0456     & .2400     & .1950 \\
  & .3600     & .0426     & .2640     & .2070 \\
  & .3694     & .0452     & .2444     & .2006 \\
  & .3551     & .0476     & .2638     & .2053 \\
  & .35506925 & .04762420 & .26376978 & .20533736
\\  \hline
2 & .1220     & .2940     & .0120     & .1510 \\
  & .1110     & .3500     & .0122     & .0989 \\
  & .1190     & .3180     & .0137     & .1200 \\
  & .1140     & .3423     & .0112     & .1036 \\
  & .1149     & .3201     & .0127     & .1157 \\
  & .11488206 & .32010981 & .01270236 & .11565949
\\ \hline
3 & .0280     & .2150     & .1510     & .0887 \\
  & .0174     & .2230     & .2100     & .0905 \\
  & .0234     & .2220     & .1740     & .0925 \\
  & .0183     & .2250     & .2031     & .0842 \\
  & .0223     & .2215     & .1728     & .0943 \\
  & .02224589 & .22154115 & .17282552 & .09426904\\
\hline
\end{tabular}
}
\end{center}
\caption{
Franck-Condon (squared) overlap  between states
involved in the transition $A^2\Pi _2 \ - \ X^2 \Sigma^+$
of CN molecule.  The parameters of this transition are:
$j_1=60.4084$, $\beta_1 = 1.2167 \ a_0^{-1}$, $R_{1} = 2.3308 \ a_0$.
$j_2=73.5602$, $\beta_2 = 1.227 \ a_0^{-1}$, $R_{2} = 2.2152 \ a_0$.
Successive entries  correspond to:
Nicholls\cite{Nicholls:81}, McCallum\cite{McCallum:70},
Waldenstrom\cite{Walden:72}, Rivas\cite{Rivas:92},
Ley-Koo\cite{Ley-Koo:95} and our results using formula (\ref{f12})
with the scaled integral (\ref{Ia}).}
\label{villa:T12}
\end{table}

\clearpage

\begin{table}[h]
\begin{center}
{\small
\begin{tabular}{|c|c|c|c|c|c|c|c|} \hline
$\nu_1 \diagdown \nu_2$ & 0    &  1  &  2   &  3   &  4  & 5 & 6   \\
\hline
0 & 3.40 -- 1 & 4.06 -- 1 & 2.00 -- 1 & 5.00 -- 2 & 6.00 -- 3 & .00000 & .00000  \\
  & 4.06 -- 1 & 4.01 -- 1 & 1.58 -- 1 & 3.17 -- 2 & 3.47 -- 3 & 2.01  --4 & 5.72 --6 \\
  & 3.38 -- 1 & 4.06 -- 1 & 1.97 -- 1 & 5.00 -- 2 & 7.00 -- 3 & .00000 & .0000 \\
  & 4.10 -- 1 & 3.98 -- 1 & 1.62 -- 1 & 3.42 -- 2 &  --       &  --     & --  \\
  & 3.99 -- 1 & 3.96 -- 1 & 1.64 -- 1 & 3.61 -- 2 &  --       &  --     & --  \\
  & 4.40 -- 1 & 3.91 -- 1 & 1.40 -- 1 & 2.59 -- 2 & 2.67 -- 3 & 1.54 -- 4 & 4.70 -- 6 \\
&.44008920 & .39144379 & .13979116 & .02585242 & .00266511 & .00015358 & .00000470 \\
\hline
1 & 3.23 -- 1 & 2.00 -- 3 & 2.12 -- 1 & 3.01 -- 1 & 1.34 -- 1 &2.70 -- 2 & 3.00 -- 3 \\
  & 3.27 -- 1 & 3.71 -- 3 & 2.85 -- 1 & 2.77 -- 1 & 9.18 -- 2 &1.41 -- 2 & 1.07 -- 3\\
  & 3.24 -- 1 & 2.00 -- 3 & 2.12 -- 1 & 2.98 -- 1 & 1.31 -- 1 &2.70 -- 2 & 3.00 -- 3\\
  & 3.31 -- 1 & 2.90 -- 3 & 2.74 -- 1 & 2.76 -- 1 &  --       & --       &  --  \\
  & 3.34 -- 1 & 2.80 -- 3 & 2.65 -- 1 & 2.77 -- 1 &  --       & --       &  --  \\
  & 3.28 -- 1 & 1.39 -- 2 & 3.11 -- 1 & 2.58 -- 1 & 7.71 -- 2 &1.11 -- 2 & 8.23 -- 4 \\
&.32752991 & .01385303 & .31130862 & .25822192 & .07714093 &.01109089 & .00082330 \\
\hline
2 &1.90 -- 1 &1.03 -- 1 &1.13 -- 1 &3.90 -- 2 &2.73 -- 1 &2.10 -- 1 &5.90 -- 2 \\
  &1.64 -- 1 &1.59 -- 1 &6.59 -- 2 &1.05 -- 1 &3.06 -- 1 &1.63 -- 1 &3.41 -- 2\\
  &1.90 -- 1 &1.03 -- 1 &1.13 -- 1 &3.90 -- 2 &2.73 -- 1 &2.10 -- 1 &6.10 -- 2\\
  &1.66 -- 1 &1.59 -- 1 &6.88 -- 2 &9.56 -- 2 &--        &--        &--     \\
  &1.70 -- 1 &1.64 -- 1 &6.60 -- 2 &8.98 -- 2 &--        &--        &--     \\
  &1.50 -- 1 &1.92 -- 1 &4.26 -- 2 &1.42 -- 1 &3.02 -- 1 &1.41 -- 1 &2.75  -- 2 \\
&.15030284 & .19173591 & .04256041 & .14151502 & .30231160 &.14140430 & .02748643 \\
\hline
3 &8.80 -- 2&1.77 -- 1 &2.00 -- 3&1.61 -- 1 &1.00 -- 3 &1.84 -- 1 &2.59 -- 1  \\
  &6.67 -- 2&1.93 -- 1 &2.25 -- 2&1.50 -- 1 &1.11 -- 2 &2.59 -- 1 &2.26 -- 1     \\
  &8.80 -- 2&1.78 -- 1 &1.00 -- 3&1.62 -- 1 &2.00 -- 3 &1.80 -- 1 &2.60 -- 1    \\
  &6.69 -- 2&1.97 -- 1 &2.21 -- 2&1.52 -- 1 &--        &--        &--    \\
  &6.63 -- 2&2.07 -- 1 &2.50 -- 2&1.42 -- 1 &--        &--        &--    \\
  &5.56 -- 2&1.99 -- 1 &4.62 -- 2&1.32 -- 1 &2.93 -- 2 &2.76 -- 1 &2.03 -- 1 \\
&.05560264 & .19857737 & .04615778 & .13209793 & .02933752 &.27596042 & .20331880 \\
\hline
4 &3.60 -- 2 &1.45 -- 1 &7.40 -- 2 &3.10 -- 2 &1.13 -- 1 &4.60 -- 2&8.30 -- 2 \\
  &2.44 -- 2 &1.29 -- 1 &1.22 -- 1 &4.67 -- 3 &1.53 -- 1 &6.94 -- 3&1.76 -- 1\\
  &3.60 -- 2 &1.45 -- 1 &7.70 -- 2 &3.20 -- 2 &1.14 -- 1 &4.80 -- 2&8.30 -- 2\\
  &--        &--        &--        &--        &--        &--       &--    \\
  &--        &--        &--        &--        &--        &--       &--    \\
  &1.84 -- 2 &1.17 -- 1 &1.49 -- 1 &1.60 -- 6 &1.60 -- 1 &2.78 -- 4&2.07 -- 1 \\
&.01835737 & .11726744 & .14897452 & .00000132 & .16030221 &.00027803 & .20656645 \\
\hline
\end{tabular}
}
\end{center}
\caption{Franck-Condon (squared) overlap between states
involved in the transition $B^3 \Pi_g \ - \ A^3 \Sigma^+_u$
of N$_2$ molecule. The parameters of this transition are:
$j_1 = 58.55$, $\beta_1 = 1.301 \ a_0^{-1}$, $R_{1} = 2.287 \ a_0$.
$j_2 = 50.994$, $\beta_2 = 1.280 \ a_0^{-1}$, $R_{2} = 2.420 \ a_0$.
Successive entries  correspond to:
Jarmain \cite{Jarmain:54}, Zare \cite{Zare:65}, Chang
\cite{Chang:70}, Waldenstrom \cite{Walden:72}, Palma \cite{Palma:92}, Ley-Koo
\cite{Ley-Koo:95} and our results using formula (\ref{f12}) with the scaled integral
(\ref{Ia}).}
\label{villa:T8}
\end{table}

\addtocounter{table}{-1}
\begin{table}[h]
\begin{center}
{\small
\begin{tabular}{|c|c|c|c|c|c|c|c|} \hline
$\nu_1 \diagdown \nu_2$ & 0    &  1  &  2   &  3   &  4  & 5 & 6   \\
\hline
5 &1.40-- 2 &8.90 -- 2 &1.26-- 1 &9.00-- 3 &8.60-- 2 &4.40-- 2 &1.07-- 1 \\
  &8.38-- 3 &6.57 -- 2 &1.39-- 1 &4.09-- 2 &4.94-- 2 &1.00-- 1 &5.05-- 2 \\
  &1.40-- 2 &8.60 -- 2 &1.27-- 1 &9.00-- 3 &8.80-- 2 &4.30-- 2 &1.04-- 1 \\
  &--       &--        &--       &--       &--       &--       &--       \\
  &--       &--        &--       &--       &--       &--       &--       \\
  &5.69-- 3 &5.37 -- 2 &1.43-- 1 &7.00-- 2 &2.74-- 2 &1.27-- 1 &2.84-- 2 \\
&.00569285 & .05367159 & .14280467 & .06997892 & .02736416 &.12678991 & .02832830 \\
\hline
6 &5.00 -- 3& 4.20 -- 2&1.19 -- 1&6.90 -- 2&4.00 -- 3&1.04 -- 1&3.00 -- 3\\
  &2.80 -- 3& 2.92 -- 2&9.94 -- 2&1.03 -- 1&2.04 -- 3&9.29 -- 2&4.02 -- 2\\
  &5.00 -- 3& 4.40 -- 2&1.13 -- 1&6.90 -- 2&5.00 -- 3&1.06 -- 1&3.00 -- 3\\
  &--       & --       &--       &--       &--       &--       &--       \\
  &--       & --       &--       &--       &--       &--       &--       \\
  &1.71 -- 3& 2.14 -- 2&8.95 -- 2&1.25 -- 1&1.45 -- 2&7.41 -- 2&6.89 -- 2\\
&.00170792 & .02138991 & .08952954 & .12450143 & .01447182 &.07414661 & .06915838 \\
\hline
\end{tabular}
}
\end{center}
\caption{Continued}
\end{table}


\begin{thebibliography}{40}

\bibitem{Morse:1929} P.M. Morse, {\it Phys. Rev. \bf 34}, 57 (1929).

\bibitem{Child:84} M.S. Child, L. Halonen,  {\it  Adv. Chem. Phys. \bf
         57}, 1 (1984).

\bibitem{Franck:AM} A. Frank and P. van Isacker, {\it Albegraic Methods in
    Molecular and Nuclear Structure}; Wiley Interscience, NY 1994.

\bibitem{Iach:JCP98}  F. Iachello, M. Ibrahim,
{\it J. Phys. Chem. \bf A 102}, 9427 (1998).

\bibitem{Muller:98} T. Muller, P. Dupr\'e, P.H. Vaccaro,
  F. Perez-Bernal, M. Ibrahim and F. Iachello 
{\it Chem. Phys. Lett. \bf 292}, 243 (1998).

\bibitem{Muller:99} T. Muller, P.H. Vaccaro, F. Perez-Bernal, F. Iachello 
{\it J. Chem. Phys. \bf 111}, 5038 (1999).

\bibitem{Muller:2000} T. Muller, P.H. Vaccaro, F. Perez-Bernal and
  F. Iachello. 
{\it Chem. Phys. Lett. \bf 329}, 271 (2000).

\bibitem{Frank:99} A. Frank, R. Lemus, M. Carvajal, C. Jung and E. Ziemniak 
{\it Chem. Phys. Lett. \bf 308}, 91 (1999).

\bibitem{Franck:25} J. Franck, {\it Trans. Faraday Soc. \bf 21}, 536
  (1925).

\bibitem{Condon:1926} E.U. Condon, {\it Phys. Rev. \bf 28}, 1182 (1926); 
                                   {\it Phys. Rev. \bf 32}, 858 (1928). 



\bibitem{Carvajal:99}  M. Carvajal, J.M. Arias and J. G\'omez-Camacho 
  {\it Phys. Rev. \bf A 59}, 3462 (1999).


\bibitem{curro:99} F. P\'erez-Bernal {\it Configuration Localized Stated
  and $1-D$ Franck-Condon Integrals}, unpublished (1999). 

\bibitem{Frank:PRA2000} A. Frank, A. L. Rivera and K. B. Wolf {\it
    Phys. Rev. \bf A 61}, 054102 (2000).


\bibitem{Ley-Koo:95} E. Ley-Koo and S. Mateos-Cortes and
  G. Villa-Torres {\it Int. J. Quantum Chem. \bf 56}, 175 (1995);
G. Villa-Torres,  Thesis, UNAM, M\'exico (1996).


\bibitem{Gradshteyn} J. M. Ryzhik, I. S. Gradshteyn  
{\it Table of Integrals, Series, and Products}, Fifth edition (1994).

\bibitem{Herzberg}  K. Huber, G. Herzberg, {\it Constants of Diatomic
    Molecules}, Van Nostrand, Reinhold, NY, 1979; 
   G. Herzberg, {\it Molecular Spectra and Molecular
    Structure. I. Spectra of Diatomic Molecules}, Van Nostrand,
    Reinhold, NY, 1950.

\bibitem{Renato:2001} R. Lemus and A. Frank, {\it (To be published)}


\bibitem{Fraser:53} P.A. Fraser and W.R. Jarmain {\it
    Proc. Phys. Soc. \bf A 66}, 1145 (1953). 


\bibitem{Berrondo:80} M. Berrondo and A. Palma, {\it J. Phys. {\bf A}
      Math. Gen. \bf 13}, 773  (1980).

\bibitem{Rivas:92} J.F. Rivas-Silva, G. Campoy and A. Palma,
{\it Int. J. Quant. Chem. \bf 43}, 747 (1992).


\bibitem{Matsumoto:93}  A. Matsumoto and  K. Iwamoto, {\it
    J. Quant. Spectrosc. Radiat. Transfer \bf 50}, 103 (1993).


\bibitem{Kusch:77} P. Kusch and M. M. Hessel {\it J. Chem. Phys. \bf
    67}, 586 (1978).

\bibitem{Hessel:79} M. M. Hessel and C.R. Vidal,  {\it
    J. Chem. Phys. \bf 70},  4439 (1979).


\bibitem{McCallum:70} J.C. McCallum, W.R. Jarmain and R.W. Nicholls 
{\it ``Franck-Condon factors and related quantities for diatomic
  molecular band systems''}; 
{\it Spectroscopic Reports \bf 1-5}, Centre for Research in
Experimental Space Science, York University (1970-72).

\bibitem{Walden:72} S. Waldenstrom and K. Razi-Naqvi,
{\it J. Chem. Phys. \bf 87}, 3563 (1987).

\bibitem{Zare:65} R.N. Zare, E.O. Larsson and R.A. Berg, {\it
    J. Mol. Spectrosc. \bf 15}, 117 (1965).

\bibitem{Drallos:86} P. J. Drallos and J. M. Wadehra 
{\it J.  Chem. Phys. \bf 85}, 6524 (1986).

\bibitem{Nicholls:81} R.W. Nicholls, {\it J. Chem. Phys. \bf 74},
  6980 (1981).

\bibitem{Palma:92} A. Palma, J.R. Rivas-Silva, J.S. Durand,
 L. Sandoval, {\it  Int. J. Quant. Chem. \bf 41}, 811 (1992).


\bibitem{Jarmain:54} W.R. Jarmain and R.W. Nicholls, 
{\it Can. J. Phys. \bf 32}, 201 (1954).


\bibitem{Chang:70} T.V. Chang and M. Karplus, {\it J. Chem. Phys. \bf
    52}, 783 (1970).


\bibitem{Bunker:98} P.R. Bunker and P. Jensen, {\it Molecular symmetry
    and spectroscopy}, NRC-CNRC, Ottawa, Canada, 1998.



\end{thebibliography}
\end{document}